\documentstyle[aps,12pt,epsf]{revtex}
\def\beq{\begin{equation}}
\def\eeq{\end{equation}}
\def\bea{\begin{eqnarray}}
\def\eea{\end{eqnarray}}

\def\vel{\left|}
\def\ver{\right|}
\def\nnb{\nonumber}
\def\ga{\left(}
\def\dr{\right)}

\def\nnb{\nonumber}

\def\ba{\begin{array}}
\def\ea{\end{array}}

\def\bos{\lower 0.5cm\hbox{{\vrule width 0pt height 1.3cm}}}
\def\aaa{\lower 0.cm\hbox{{\vrule width 0pt height .8cm}}}
\def\dol{\lower 0.6cm\hbox{{\vrule width 0pt height .8cm}}}
\begin{document}

\bigskip
{\Large\bf
\centerline{$B\rightarrow X_s\tau^+\tau^-$ 
in a CP softly
broken}  
\centerline{two Higgs doublet model}
\bigskip
\normalsize

\centerline{Chao-Shang Huang$^1$,~~~~Shou Hua Zhu$^{2,1}$}
\centerline{\sl $^1$ Institute of Theoretical Physics, Academia Sinica,
      P.O.Box 2735,}
\centerline{\sl Beijing 100080, P.R.China}
\centerline{\sl $^2$ CCAST (World Laboratory), Beijing 100080, P.R.China} 
\bigskip

\begin{abstract}
 The differential branching ratio, forward-backward asymmetry,
CP asymmetry and 
lepton polarization
for a B-meson to decay to strange hadronic final states and a
$\tau^+\tau^-$ pair in a CP softly broken 
two Higgs doublet model are computed. 
It is shown that
contributions of neutral Higgs bosons  to the decay are 
quite significant when $\tan\beta$ is large.
And it is proposed to measure the direct 
CP asymmetry in back-forward asymmetry.
\end{abstract}

\vfill\eject\pagestyle{plain}\setcounter{page}{1}

\section{Introduction}

The origin of the CP violation has been one of main issues
in high energy physics since the discovery of the CP violation in 
the $K_0-\overline{K}_0$ systerm in 1964 \cite{new1}.
The measurements of electric dipole moments of the neutron and 
electron and the matter-antimatter asymmetry in the universe
indicate that one needs new sources of CP violation in addition
to the CP violation come from CKM matrix, which has been one 
of motivations to search new theoretical models beyond the 
standard model (SM).

The minimal extension of the SM is to enlarge the Higgs sectors
of the SM \cite{new2}. It has been shown that if one adheres to the
natural flavor conservation (NFC) in the Higgs sector, then
a minimum of three Higgs doublets are necessary in order to have
spontaneous CP violations \cite{new3}. However, the constraint
can be evaded if one allows the real and image parts of $\phi_1^+ 
\phi_2$ have different self-couplings and adds a linear term of Re($\phi^+_1\phi_2$)
in the Higgs potential (see below Eq. (\ref{eq2}) with $m_4^2=0$).
Then, one can construct a CP spontaneously broken two Higgs 
doublet (2HDM), which is the minimal and the most "economical" one \footnote{ 
Comparing the Model III  2HDM \cite{new4}, in which  CP is explicitly
violated, the CP spontaneously broken 2HDM has only two new
parameters besides the masses of the Higgs bosons in the
large $\tan\beta$ limit (see below). In this sense it is the most
"economical".
}
among the extensions of the SM that provide  new source of CP violation. Furthermore,
in addition to the above terms, if one adds a linear term of Im($\phi^+_1\phi_2$), then
one has a CP softly broken 2HDM~\cite{Georgi}.

Flavor changing neutral current (FCNC) transitions $B\rightarrow X_s\gamma$ and
$B\rightarrow X_sl^+l^-$ provide testing grounds for the SM at
the loop level and sensitivity to new physics. 
Rare decays $B\rightarrow X_sl^+l^-(l=e,\mu)$ have been 
extensively investigated in both SM and the beyond 
 \cite{gsw,ex1}. 
In these processes contributions from exchanging neutral
Higgs bosons (NHB) 
can be safely neglected because of smallness of $\frac{m_l}{m_W}
(l=e,\mu)$. 
The inclusive decay $B\rightarrow X_s\tau^+
\tau^-$ has also been investigated in the SM, the model II 2HDM and
SUSY models  with and without
including the contributions of NHB \cite{new7,new8,Dai,add2}.
In this note we investigate the inclusive decay $B\rightarrow X_s\tau^+\tau^-$
 with emphasis on CP violation effect in a CP softly broken 
2HDM, which we shall call Model IV hereafter for the sake of simplisity. 
We consider the Model IV in which the up-type quarks get masses from
Yukawa couplings to the one Higgs doublet $H_2$ 
and down-type quarks and leptons get masses from Yukawa
couplings to the another Higgs doublet $H_1$.
The Higgs boson couplings to down-type quarks and leptons depend on 
only the CP violated phase $\xi$ which comes from the expectation
value of Higgs and the ratio $tg\beta
=\frac{v_2}{v_1}$ in the large $tg\beta$ limit (see next section), 
which are the free parameters
in the model. 
Because the couplings of the charged Higgs to fermions in Model IV are the
same as those in the model II, the constraints on $\tan\beta$ due to
effects arising from the charged Higgs are the same as those in the
model II.
Constraints on
$tg\beta$ from $K-\bar{K}$ and $B-\bar{B}$ mixing, $\Gamma(b\rightarrow s\gamma)
$,$\Gamma(b\rightarrow c\tau\bar{\nu}_{\tau})$ and $R_b$ have been given 
\cite{15}
\begin{equation}
0.7\le tg\beta\le 0.52(\frac{m_{H^{\pm}}}{1Gev})
\end{equation}
(and the lower limit $m_{H^{\pm}}\ge 200Gev$ has also been given in the ref.
\cite{15}). It is obvious that the contributions from exchanging neutral Higgs
bosons now is enhanced roughly by a factor of $tg^2\beta$ and can compete with
those from exchanging $\gamma,~Z$ when $tg\beta$ is large enough.
 Because the CP violation effects in $B\rightarrow X_s \tau^+\tau^-$ come
 from the couplings of NHB to leptons and quarks,
 we shall be interested in the large $\tan\beta$ limit in this note.
 The constraints on $\xi$ can be obtained from the electric dipole moments
 (EDM) of the neutron and electron, which will be analysed in the next
 section.

\section{Model description}
Consider two complex $y=1$, $SU(2)_w$ doublet scalar fields, $\phi_1$ and
$\phi_2$. The Higgs potential which spontaneously breaks $SU(2)\times U(1)$
down to $U(1)_{EM}$ can be written in the following form \cite{Georgi}:
\begin{eqnarray}
V(\phi_1,\phi_2) &=&
\sum_{i=1,2} [m_i^2 \phi_i^+ \phi_i +\lambda_i  (\phi_i^+ \phi_i)^2] \nonumber\\
&&+ m_3^2 Re(\phi_1^+\phi_2) + m_4^2 Im(\phi_1^+ \phi_2) \nonumber\\
&&+ \lambda_3 [ (\phi_1^+ \phi_1)(\phi_2^+ \phi_2) ]
+ \lambda_4 [ (\phi_1^+ \phi_2) (\phi_2^+ \phi_1) ]
\nonumber \\
&&
+ \lambda_5
 [ \mbox{Re}(\phi_1^+ \phi_2)]^2
+ \lambda_6
 [ \mbox{Im}(\phi_1^+ \phi_2) ]^2
 \label{eq2}
\end{eqnarray}
Hermiticity requires that all parameters are real. 
The potential is CP softly broken due to the presence of the term $m_4^2 Im(\phi_1^+ \phi_2)$.
It is easy to see that the minimum of the potential is at
\begin{eqnarray}
<\phi_1>=\left( \begin{array}{c}
0 \\
v_1
\end{array}
\right), \ \ \ \
<\phi_2>=\left( \begin{array}{c}
0 \\
v_2 e^{i\xi}
\end{array}
\right),
\end{eqnarray}
thus breaking $SU(2)\times U(1)$
down to $U(1)_{EM}$ and
simutaneously breaking CP, as desired. 
It should be noticed that only for  $\lambda_5 \neq \lambda_6$, the phase
$\xi$ can't rotated away as usual, which breaks the CP-conservation. If $m_4^2$=0 in (2)
then the potential is CP invarint. It has been shown that the CP spontaneously breaking
happens at (3)~\cite{leb}. We limit ourself to the case of $m_4^2 \neq 0$ in the paper
and shall investigate the $m_4^2$=0 case in a separate paper ~\cite{hz}.

In the following we will work out the mass spectrum of the Higgs
boson. For charged components, the mass-squared matrix for negtive states is
\begin{eqnarray}
\lambda_4 \left(
\begin{array}{cc}
v_1^2 & -v_1 v_2 e^{i\xi} \\
-v_1 v_2 e^{-i\xi} & v_2^2
\end{array}
\right),
\end{eqnarray}
Diagonalizing the mass-squared matrix results in one 
zero-mass Goldstone state:
\begin{eqnarray}
G^-=e^{i\xi} \sin\beta \phi_2^- +\cos\beta \phi_1^-,
\end{eqnarray}
and one massive charged Higgs boson state:
\begin{eqnarray}
H^-=e^{i\xi} \cos\beta \phi_2^- - \sin \beta \phi_1^-,
\\
m_{H^-}=|\lambda_4| (v_1^2+v_2^2),
\end{eqnarray}
where $\tan\beta = v_2/v_1$.
Correspondingly we could also get the positive states $G^+$ and $H^+$ with the
same masses zero and $|\lambda_4| (v_1^2+v_2^2)$, respectively.
 
For neutral Higgs components, because CP-conservation is breaking, the 
mass-squared matrix is $4\times 4$, which could not be simply 
separated into
two $2\times 2$ matrices as usual. 
However, in
the case of large $\tan\beta$ which is we intrested in,
the neutral parts can be
written as separately two $2\times 2$ matrices and one of them is
\begin{eqnarray}
v_2^2 \left(
\begin{array}{cc}
\frac{\lambda_5+ \lambda_6+(\lambda_6-\lambda_5) \cos(2\xi)}{2} &
- \frac{(\lambda_6-\lambda_5) \sin(2\xi)}{2} \\
- \frac{(\lambda_6-\lambda_5) \sin(2\xi)}{2} &
\frac{\lambda_5+ \lambda_6+(\lambda_5-\lambda_6) \cos(2\xi)}{2}
\end{array}
\right).
\end{eqnarray}
Diagonalizing the Higgs boson mass-squared matrix results in two
eigenstates:
\begin{eqnarray}
\left(
\begin{array}{c}
H^0_1\\
H^0_2
\end{array}
\right) =
\sqrt{2} \left(
\begin{array}{cc}
c_\xi & -s_\xi\\
s_\xi & c_\xi
\end{array}
\right)
\left(
\begin{array}{c}
{\rm Im} \phi_1^0 \\
{\rm Re} \phi_1^0
\end{array}
\right)
\label{eq9}
\end{eqnarray}
with masses
\begin{eqnarray}
m_{H^0_1}^2=\lambda_5 v_2^2 \nonumber \\
m_{H^0_2}^2=\lambda_6 v_2^2, 
\label{eq10}
\end{eqnarray}
where $c_\xi=\cos\xi$ and $s_\xi=\sin\xi$. The diagonalizing of the
$4\times 4$ neutral Higgs mass-squared matrix has been analytically 
carried out under some assumptions in Ref. \cite{Vend} and the results
reduce to Eq. (\ref{eq9}) and (\ref{eq10}) in the case of large
$\tan\beta$.

The another $2\times 2$ matrix can be similarly deal with.
Because the couplings of the third physical neutral Higgs boson 
and neutral Goldstone to down-type quarks and leptons 
are not enhanced for large $\tan\beta$ case in which we are interested,
we do not show the explicit results.

Now, we turn to the discussion of the Higgs-fermion-fermion couplings. 
After completing the transformation from the weak states to the mass
states, the couplings of neutral Higgs to fermions which are 
relevant to our analysis are
\begin{eqnarray}
H^0_1 \bar f f:\ \ \ \ &&  \frac{i g m_f}{2 m_w \cos\beta} 
(s_\xi- i c_\xi \gamma_5) 
\nonumber\\
H^0_2 \bar f f: \ \ \ \
 &&  -\frac{i g m_f}{ 2 m_w \cos\beta} (c_\xi+ i s_\xi \gamma_5) 
 \label{eq11}
\end{eqnarray}
where $f$ represents down-type quarks and leptons. And the
couplings of the charged Higgs bosons to fermions are the same as
those in the
CP-conservative 2HDM (model II, for examples see Ref. \cite{19}).
This is in contrary with the model III in which the couplings
of the charged Higgs to fermions are quite different from model II.
It is easy to see from Eq. (\ref{eq11}) that the contributions
come from exchanging NHB is proportional to 
$\sqrt{2} G_F s_\xi c_\xi  m_f^2/\cos^2\beta$, so that the
constaints due to EDM translate into the constraints
on $\sin 2\xi \tan^2\beta$ ($1/\cos\beta \sim \tan\beta$ in the large 
$\tan\beta$ limit). According to the analysis in Ref. \cite{new6},
we have the constraint
\begin{eqnarray}
\sqrt{|\sin 2\xi|}\tan\beta < 50
\label{eqa}
\end{eqnarray}
from the neutron EDM. And the constraint from the electron EDM
is not stronger than Eq. (\ref{eqa}). It is obvious from Eq. (\ref{eqa}) that there
is a constraint on $\xi$ only if $\tan \beta >  50$ and the stringent constraint on $\tan \beta$
comes out and is $\tan \beta < 50$ when $\xi = \pi $/4.

\section{Formula for $B \rightarrow X_{s} \tau^{+} \tau^{-}$ }
Inclusive decay rates of heavy hadrons can be calculated in heavy quark 
effective theory (HQET) \cite{17} and it has been shown that the 
leading terms in $1/m_Q$
expansion turn out to be the decay of a free (heavy) quark and corrections stem
from the order $1/m_Q^2$ \cite{18}. In what follows we shall 
calculate the leading term.
The transition rate for
$b\rightarrow s\tau^+\tau^-$ can be computed in the framework of the QCD 
corrected effective weak hamiltonian, obtained by integrating out the top quark,
Higgs bosons and $W^{\pm},Z$ bosons
\begin{equation}\label{ham}
H_{eff}=\frac{4G_F}{\sqrt{2}}V_{tb}V^*_{ts}(\sum_{i=1}^{10}C_i(\mu)O_i(\mu)
+\sum_{i=1}^{10}C_{Q_i}(\mu)Q_i(\mu))
\end{equation}
where $O_i(i=1,\cdots ,10)$ is the same as that given in the ref.\cite{gsw}, $Q_i$'s
come from exchanging the neutral Higgs bosons and are defined in Ref. 
\cite{Dai}. The explicit expressions of the operators governing $B\rightarrow X_s\tau^+\tau^-$ are given 
as follows:
\begin{eqnarray}
O_7 &=& (e/16\pi^2) m_b (\bar{s}_{L\alpha} \sigma^{\mu\nu}
b_{R\alpha}) F_{\mu\nu}, \nonumber \\
O_8 &=& (e/16\pi^2) (\bar{s}_{L\alpha} \gamma^{\mu}
b_{L\alpha}) \bar\tau \gamma_{\mu} \tau, \nonumber \\
O_9 &=& (e/16\pi^2) (\bar{s}_{L\alpha} \gamma^{\mu}
b_{L\alpha}) \bar\tau \gamma_{\mu} \gamma_5 \tau, \nonumber \\
Q_1 &=& (e^2/16\pi^2) (\bar{s}_{L\alpha} b_{R\alpha})
 (\bar\tau \tau), \nonumber \\
Q_2 &=& (e^2/16\pi^2) (\bar{s}_{L\alpha} b_{R\alpha})
 (\bar\tau \gamma_5 \tau).
\end{eqnarray}

At the renormalization point $\mu=m_W$ the coefficients $C_i$'s in the 
effective hamiltonian have been given in the ref.\cite{gsw} and $C_{Q_i}$'s are
(neglecting the $O(tg\beta)$ term)
\begin{eqnarray}
C_{Q_1}(m_W)&=&\frac{m_bm_{\tau}tg^2\beta x_t}{2 sin^2\theta_W}
\{
\sum_{i=H_1,H_2} \frac{ A_{i}}{m_{i}^2} (f_1 B_{i}+f_2 E_i) \},
\nonumber \\
C_{Q_2}(m_W)&=&\frac{m_bm_{\tau}tg^2\beta x_t}{2 sin^2\theta_W}
\{
\sum_{i=H_1,H_2} \frac{ D_{i}}{m_{i}^2} (f_1 B_{i}+f_2 E_i) \},
\nonumber \\
C_{Q_3}(m_W)&=&\frac{m_be^2}{m_{\tau}g_s^2}(C_{Q_1}(m_W)+C_{Q_2}(m_W)),
\nonumber 
\\
C_{Q_4}(m_W)&=&\frac{m_be^2}{m_{\tau}g_s^2}(C_{Q_1}(m_W)-C_{Q_2}(m_W)), \nonumber
\\
C_{Q_i}(m_W)&=&0, ~~~~i=5,\cdots, 10
\label{eq1}
\end{eqnarray}
where
\begin{eqnarray}
A_{H_1}&=-s_\xi,  \  D_{H_1}&=i c_\xi,
\nonumber \\
A_{H_2}&= c_\xi, \  D_{H_2}&=i s_\xi,
\nonumber \\
B_{H_1}&=\frac{ i c_\xi-s_\xi}{2}, \
B_{H_2}&=\frac{c_\xi+i s_\xi}{2}, 
\nonumber 
\end{eqnarray}
\begin{eqnarray} 
f_1&=& \frac{x_t ln x_t}{x_t-1}-
\frac{x_{H^\pm} ln x_{H^\pm}-x_t ln x_t }{x_{H^\pm}-x_t},
\nonumber \\
f_2&=& \frac{x_t ln x_t}{(x_t-1)(x_{H^\pm}-1)}-
\frac{x_{H^\pm} ln x_{H^\pm} }{(x_{H^\pm}-x_t)(x_{H^\pm}-1)}
\end{eqnarray}
with $ x_i=m_i^2/m_w^2$.
In Eq. (\ref{eq1}), $E_i$ are given by
\begin{eqnarray} 
E_{H_1}&=& \frac{1}{2} (-s_\xi c_1+ c_\xi c_2), 
\nonumber \\
E_{H_2}&=& \frac{1}{2} (c_\xi c_1+ s_\xi c_2), 
\nonumber \\
c_1&=& -x_{H^\pm}+c_\xi x_{H_1} (c_\xi+i s_\xi)
+s_\xi x_{H_2} (s_\xi-i c_\xi),
\nonumber \\
c_2&=& i\left(- x_{H^\pm}+s_\xi x_{H_1} (s_\xi-i c_\xi)
+c_\xi x_{H_2} (c_\xi+i s_\xi) \right).
\end{eqnarray}

Neglecting the strange quark mass, the effective hamiltonian (\ref{ham}) leads 
to the following matrix element for $b\rightarrow s\tau^+\tau^-$
\begin{eqnarray}\label{matrix}\nonumber
M&=&\frac{G_F\alpha}{\sqrt{2}\pi}V_{tb}V^*_{ts}[C^{eff}_8\bar{s}_L\gamma_{\mu}
b_L\bar{\tau}\gamma^{\mu}\tau+C_9\bar{s}_L\gamma_{\mu}b_L\bar{\tau}\gamma^{\mu}
\gamma^5\tau\\
&+&2C_7m_b\bar{s}_Li\sigma^{\mu\nu}\frac{q^{\nu}}{q^2}b_R\bar{\tau}\gamma^{\mu}
\tau+C_{Q_1}\bar{s}_Lb_R\bar{\tau}\tau+C_{Q_2}\bar{s}_Lb_R\bar{\tau}\gamma^5
\tau],
\end{eqnarray}
where \cite{gsw,new7,10}
\begin{eqnarray}\label{coeff}\nonumber
C^{eff}_8&=&C_8+\{g(\frac{m_c}{m_b},\hat{s})\\
&+&\frac{3}{\alpha^2}k\sum_{V_i=
\psi^{\prime}, \psi^{\prime\prime}...}
\frac{\pi M_{V_i}\Gamma(V_i\rightarrow\tau^+\tau^-)}{M^2_{V_i}-q^2
-iM_{V_i}\Gamma_{V_i}}\}(3C_1+C_2),
\end{eqnarray}
with $\hat{s}=q^2/m_b^2,~~q=(p_{\tau^+}+p_{\tau^-})^2$. In (\ref{coeff}) 
$g(\frac{m_c}{m_b},\hat{s})$ arises from the one-loop matrix element 
of the four-quark 
operators and can be found in Refs. \cite{gsw,dba}.
 The second term 
in braces in 
(\ref{coeff})
estimates the long-distance contribution from the intermediate, 
$\psi^{\prime}$,
$\psi^{\prime\prime}$ ...
\cite{gsw,10}. In our numerical calculations, we choose 
$ k (3C_1+C_2)=-0.875$ \cite{PDG}.

The QCD corrections to coefficients $C_i$ and $C_{Q_i}$ can be incooperated
in the standard way by using the renormalization group equations. 
Although the $C_i$ at the scale $\mu=O(m_b)$ 
have been given in the next-to-leading order 
approximation (NLO) and without including mixing with $Q_i$, we
use the values of $C_i$ only in the leading order approximation (LO) 
since no $C_{Q_i}$ have been calculated in NLO.
The $C_i$ and $C_{Q_i}$ with LO QCD corrections have been
given in Ref. \cite{Dai}. 
\begin{eqnarray}\label{c7}
C_7(m_b)&=&\eta^{-16/23}\left[ C_7(m_W)-[\frac{58}{135}(\eta^{10/23}-1)
+\frac{29}{189}(\eta^{28/23}-1)]C_2(m_W) \right.
\nonumber \\
&&\left. -0.012C_{Q_3}(m_W)\right],
\label{eq18}
\end{eqnarray}
\begin{eqnarray}
C_8(m_b)&=&C_8(m_W)+\frac{4\pi}{\alpha_s(m_W)}[-\frac{4}{33}(1-\eta^{-11/23})
+\frac{8}{87}(1-\eta^{-29/23})]C_2(m_W),\\
C_9(m_b)&=&C_9(m_W),\\
C_{Q_i}(m_b)&=&\eta^{-\gamma_Q/\beta_0}C_{Q_i}(m_W),~~i=1,2,
\end{eqnarray}
where $\gamma_Q=-4$ \cite{21} is the anomalous dimension of $\bar{s}_Lb_R$,
$\beta_0=11-2 n_f/3$, and $\eta=\alpha_s(m_b)/\alpha_s(m_W)$.

After a straightforward calculation,  we obtain the 
invariant dilepton mass distribution \cite{Dai}
\begin{eqnarray}
\frac{{\rm d}\Gamma(B\rightarrow X_s\tau^{+}\tau^{-})}{{\rm d}s}
 &=& B(B\rightarrow X_c l {\bar \nu}) \frac{{\alpha}^2}
 {4 \pi^2 f(m_c/m_b)} (1-s)^2(1-\frac{4t^2}{s})^{1/2}
 \frac{|V_{tb}V_{ts}^{*}|^2}{|V_{cb}|^2} D(s) \nonumber \\
 D(s) &=& |C_8^{eff}|^2(1+\frac{2t^2}{s})(1+2s)
      + 4|C_7|^2(1+ \frac{2t^2}{s})(1+\frac{2}{s}) \nonumber \\
    & &  + |C_9|^2 [ ( 1 + 2s) + \frac{2t^2}{s}(1-4s)]
      +12 {\rm Re}(C_7 C_{8}^{eff*})(1+\frac{2t^2}{s}) \nonumber \\
  & & + \frac{3}{2}|C_{Q_1}|^2 (s-4t^2) + \frac{3}{2}|C_{Q_2}|^2s
      + 6{\rm Re}(C_9 C_{Q_2}^{*}) t
\label{eq22}
\end{eqnarray}
where s=$q^2/m_b^2$, t=$m_{\tau}/m_{b}$, 
$B(B\rightarrow X_c l {\bar \nu})$ is the branching ratio,
$f$ is the phase-space factor and f(x)=$1-8 x^2+8 x^6
-x^8-24 x^4 \ln ~ x$.

 The CP asymmetry for the $B \rightarrow X_s l^+ l^-$ and
$\overline{ B} \rightarrow \overline{ X}_s l^+ l^-$ is defined as
\begin{eqnarray}
A_{CP}^1(s)=\frac{{\rm d}\Gamma/{\rm d}s -  {\rm d}\overline{\Gamma}/{\rm d}s}{
{\rm d}\Gamma/{\rm d}s +  {\rm d}\overline{\Gamma}/{\rm d}s}.
\end{eqnarray}
We also give the forward-backward asymmetry
\begin{eqnarray}
A(s)=\frac{\int^{1}_{0}dz \frac{d^2\Gamma}{ds dz} - 
\int^{0}_{-1}dz \frac{d^2\Gamma}{ds dz}}{
\int^{1}_{0}dz \frac{d^2\Gamma}{ds dz} + 
\int^{0}_{-1}dz \frac{d^2\Gamma}{ds dz}}
=\frac{E(s)}{D(s)}
\end{eqnarray}
where $z=\cos\theta$ and $\theta$ is the angle between the momentum
of the B-meson and that of $l^+$ in the center of mass frame of the 
dileptons $\tau^+\tau^-$. Here,
\begin{eqnarray}
E(s)={\rm Re} (C_8^{eff} C_9^* s+2 C_7 C_9^*+
C_8^{eff} C_{Q1}^* t+ 2 C_7 C_{Q2}^* t).
\label{eq26}
\end{eqnarray}
The CP asymmetry in the forward-backward asymmetry 
for $B \rightarrow X_s \tau^+ \tau^-$ and
$\overline{ B} \rightarrow \overline{X}_s \tau^+ \tau^-$
is defined as
\begin{eqnarray}
A_{CP}^2(s)=
\frac{ A(s)- \overline{A} (s)}{ A(s)+ \overline{A}(s)}.
\end{eqnarray}
It is easy to see from Eq. (\ref{eq22})  that  the CP asymmetry
$A_{CP}^1$ is very small because the weak phase difference in
$C_7 C_8^{eff}$ arises from the small mixing of $O_7$ with
$Q_3$ (see Eq. (\ref{eq18})). In contrast with it, $A_{CP}^2$ 
can reach a large value when $\tan\beta$ is large, as can be seen
from Eq. (\ref{eq26}) and (\ref{eq1}).
Therefore, we propose to measure $A_{CP}^2$ in order to search for
new CP violation sources.

Let us now discuss the lepton polarization effects. We define three
orthogonal unit vectors:
\bea
\vec{e}_L &=& \frac{\vec{p}_1}{\vel \vec{p}_1 \ver}~, \nnb \\
\vec{e}_N&=& \frac{\vec{p}_{s} \times \vec{p}_1}
{\vel \vec{p}_{s} \times \vec{p}_1 \ver}~, \nnb \\
\vec{e}_T &=& \vec{e}_N \times \vec{e}_L~, \nnb
\eea
where $\vec{p}_1$ and $\vec{p}_{s}$ are the three momenta of the
$\ell^-$ lepton
and the $s$ quark, respectively, in the center of mass of the
$\ell^+~\ell^-$ system. The differential decay rate for any given spin
direction $\vec{n}$ of the $\ell^-$ lepton, where $\vec{n}$ is a unit vector  
in the $\ell^-$ lepton rest frame, can be written as
\bea
\frac{d \Gamma \ga \vec{n} \dr}{{\rm d} s} = 
\frac{1}{2} \ga \frac{d \Gamma}{{\rm d} s} \dr_{\!\!\! 0} 
\Big[ 1 + \ga P_L\, \vec{e}_L + P_N\, \vec{e}_N + P_T\, \vec{e}_T \dr \cdot
\vec{n} \Big]~,
\label{eq27}
\eea
where the subscript "0" corresponds to the unpolarized case, and $P_L,~P_T$,
and $P_N$, which correspond to the longitudinal, transverse and normal
projections of the lepton spin, respectively, are functions of $s$.
From Eq. (\ref{eq27}), one has
\bea
P_i (s) = \frac{ {\displaystyle{\frac{d \Gamma}{d s}
\ga \vec{n}=\vec{e}_i \dr -
\frac{d \Gamma}{ds}\ga \vec{n}=-\vec{e}_i \dr}} }
{ {\displaystyle{\frac{d \Gamma}{ds}\ga \vec{n}=\vec{e}_i \dr +
\frac{d \Gamma}{ds}\ga \vec{n}=-\vec{e}_i \dr}} } ~.
\eea

The calculations for the $P_i$'s (i = $L,~T,~N$) lead to the following 
results:
\bea
P_L &=&  (1-\frac{4 t^2}{s})^{1/2} \frac{D_L(s)}{D(s)},
\nonumber \\
P_N&=& \frac{3 \pi}{4 s^{1/2}} (1-\frac{4 t^2}{s})^{1/2}
\frac{D_N(s)}{D(s)},
\nonumber \\
P_T&=& -\frac{3 \pi t}{2 s^{1/2}}
\frac{D_T(s)}{D(s)},
\eea
where
\bea
D_L(s) &=& {\rm Re}\left(
 2 (1+2 s) C_8^{eff} C_9^*+12 C_7 C_9^*- 6 t C_{Q_1} C_9^*-
3 s C_{Q_1} C_{Q_2}^* \right), \nonumber \\
D_N(s) &=&  {\rm Im} \left(
2 s  C_{Q_1} C_7^*+s C_{Q_1} C_8^{eff *}+s C_{Q_2} C_9^*+
4 t C_9 C_7^*+ 2 t s C_8^{eff\ *} C_9 \right),
 \nonumber \\
D_T(s) &=&   {\rm Re}\left(
-2 C_7 C_9^*+ 4 C_8^{eff} C_7^* +\frac{4}{s} |C_7|^2-
C_8^{eff} C_9^* \right. \nonumber \\
&& \left. +s |C_8^{eff}|^2 -\frac{s-4 t^2}{2 t} C_{Q_1} C_9^*
-\frac{s}{t} C_{Q_2} C_7^*-\frac{s}{2 t} C_8^{eff} C_{Q_2}^* \right).
\label{eq30}
\eea
$P_i$ (i=L, T, N) have been given 
in the ref. [9], where there are some errors in $P_T$ and they gave only
two terms in $D_N$, the numerator of $P_N$.
We remind that $P_N$ is the CP-violating projection of the lepton spin
onto the normal of the decay plane. Because $P_N$ in $B \rightarrow
X_s l^+ l^-$ comes from both the quark and lepton sectors,
purely hadronic and leptonic CP-violating observables,
such as $d_n$ or $d_e$, do not necessarily strongly constrain $P_N$
\cite{add1}. So it is advantageous to use $P_N$ to investigate
CP violation effects in some extensions of SM \cite{add3}.
In the model IV 2HDM, as pointed out above, $d_n$ and $d_e$ constrain
$\sqrt{|\sin 2\xi|}\tan\beta$ and consequently $P_N$ through
$C_{Q_i}$ ($i=1,2$) (see Eq. (\ref{eq30})).

\section{numerical results}
The following parameters have been used in the numerical calculations:
$$
m_t=175Gev,~m_b=5.0Gev,~m_c=1.6Gev,~m_{\tau}=1.77Gev,~\eta=1.724,
$$
$$
m_{H_1}=100 Gev, ~m_{H_2}=m_{H^\pm}=200 Gev.
$$

Numerical results are shown in Figs. 1-9. From Figs. 1 and 2, we can see
that the contributions of NHB to the differential branching ratio
$d\Gamma/ds$ are significant when $\tan\beta$ is not smaller than 30 and the
masses of NHB are in the reasonable region, and the forward-backward asymmetry
$A(s)$ is more sensitive to $\tan\beta$
than $d\Gamma/ds$, which is similar to the case of the normal
2HDM without CP violation \cite{Dai}. 

The direct CP violation $A_{CP}^i$ ($i=1,2$) and CP-violating polarization 
$P_N$ of $B \rightarrow X_s \tau^+\tau^-$ are presented in Figs. 3-7, 
respectively. As expected, $A_{CP}^1$ is about $0.1$\% and hard to be 
measured. However, $A_{CP}^2$ can reach about $10$\%.
$A_{CP}^2$ is strongly dependent of the CP violation
phase $\xi$ and comes mainly from exchanging NHBs as expected.
From Figs. 6 and 7, one can see that $P_N$ is also strongly dependent of
the CP violation phase $\xi$ and can be as large as 5\% for some
values of $\xi$, which should be within the luminosity reach of
coming B factories, and comes mainly from NHB contributions in the most of range
of $\xi$.

Figs. 8 and 9 show the longitudinal and transverse polarizations 
respectively. It is obviously that the contributions of NHB can change
the polarization greatly, especially when $\tan\beta$ is large,
 and the dependence of $P_L$ on CP violation phase $\xi$ is not significant
in the most of range of $\xi$. 
The longitudinal  polarization of $B \rightarrow X_s \tau^+\tau^-$
has been calculated in SM and several new physics scenarios
\cite{new7}. Switching off the NHB contributions, our results
are in agreement with those in Ref. \cite{new7}.

In summary, we have calculated the differential braching ratio,
back-forward asymmetry, lepton polarizations and some CP violated 
observables for $B\rightarrow X_s \tau^+\tau^-$ in the model IV 2HDM.
As the main features of the model, NHB play an important role
in inducing CP violations,
in particular, for large $\tan\beta$.
We propose to measure $A_{CP}^2$, the direct CP asymmetry in back-forward
asymmetry, in stead of $A_{CP}^1$, the usual direct CP violation in
branching ratio, because the former could be observed if $\tan\beta$
is large enough (say, $\geq 30$) and the latter is too small to be 
observed. It is possible to discriminate the
model IV from the other 2HDMs by measuring the CP-violated observables
such as $A_{CP}^2$, $P_N$ if the nature chooses large $\tan\beta$.

\section*{Acknowledgments}
This research was supported in part by the National Nature Science
Foundation of China and the post doctoral foundation
of China. S.H. Zhu gratefully acknowledges useful discussions
with Wei Liao, Qi-su Yan and the
support of  K.C. Wong Education Foundation, Hong Kong.

\vfill\eject



\begin{figure}
\epsfxsize=18 cm
\centerline{\epsffile{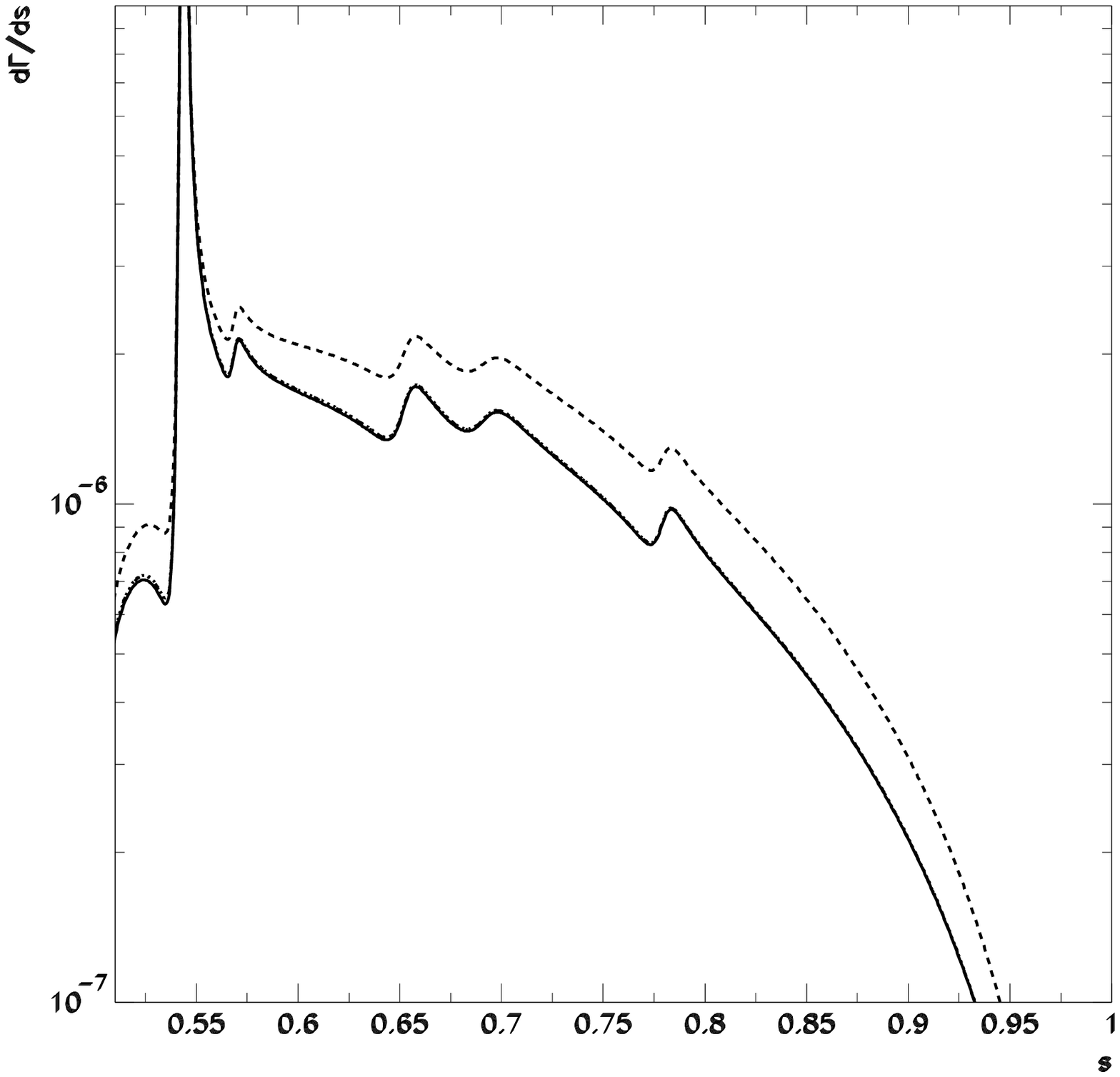}}
\caption[]{
Differential branching ratio as function of $s$,
where $\xi=\pi/4$,  solid and dashed lines represent
$\tan\beta=10$ and $30$, dot-dashed line represents the case of
switching off $C_{Q_i}$ contributions.}
\end{figure}

\begin{figure}
\epsfxsize=18 cm
\centerline{\epsffile{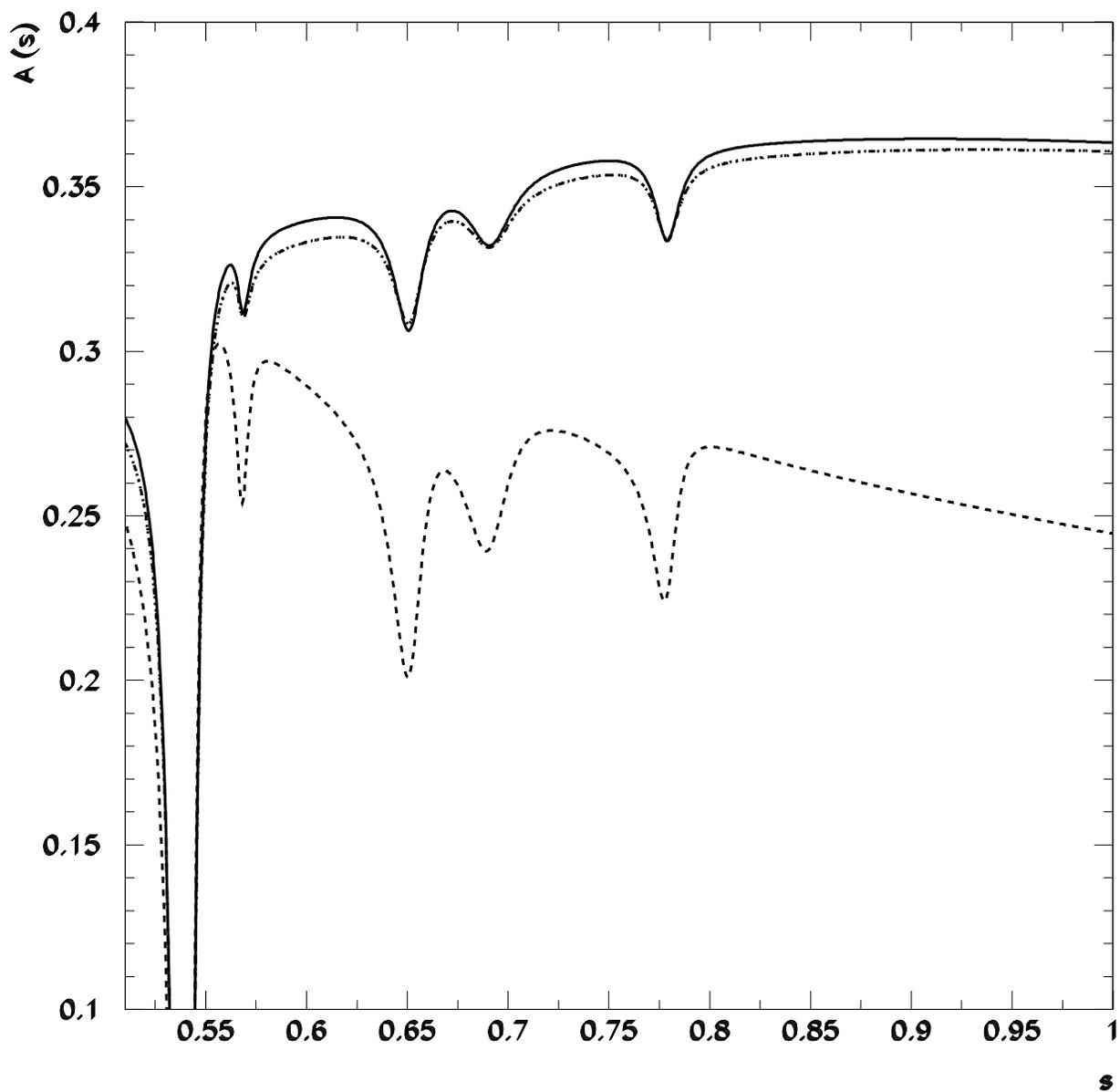}}
\caption[]{
Backward-forward asymmetry as function of $s$,
where $\xi=\pi/4$,  solid and dashed lines represent
$\tan\beta=10$ and $30$, dot-dashed line represents the case of
switching off $C_{Q_i}$ contributions.}
\end{figure}

\begin{figure}
\epsfxsize=18 cm
\centerline{\epsffile{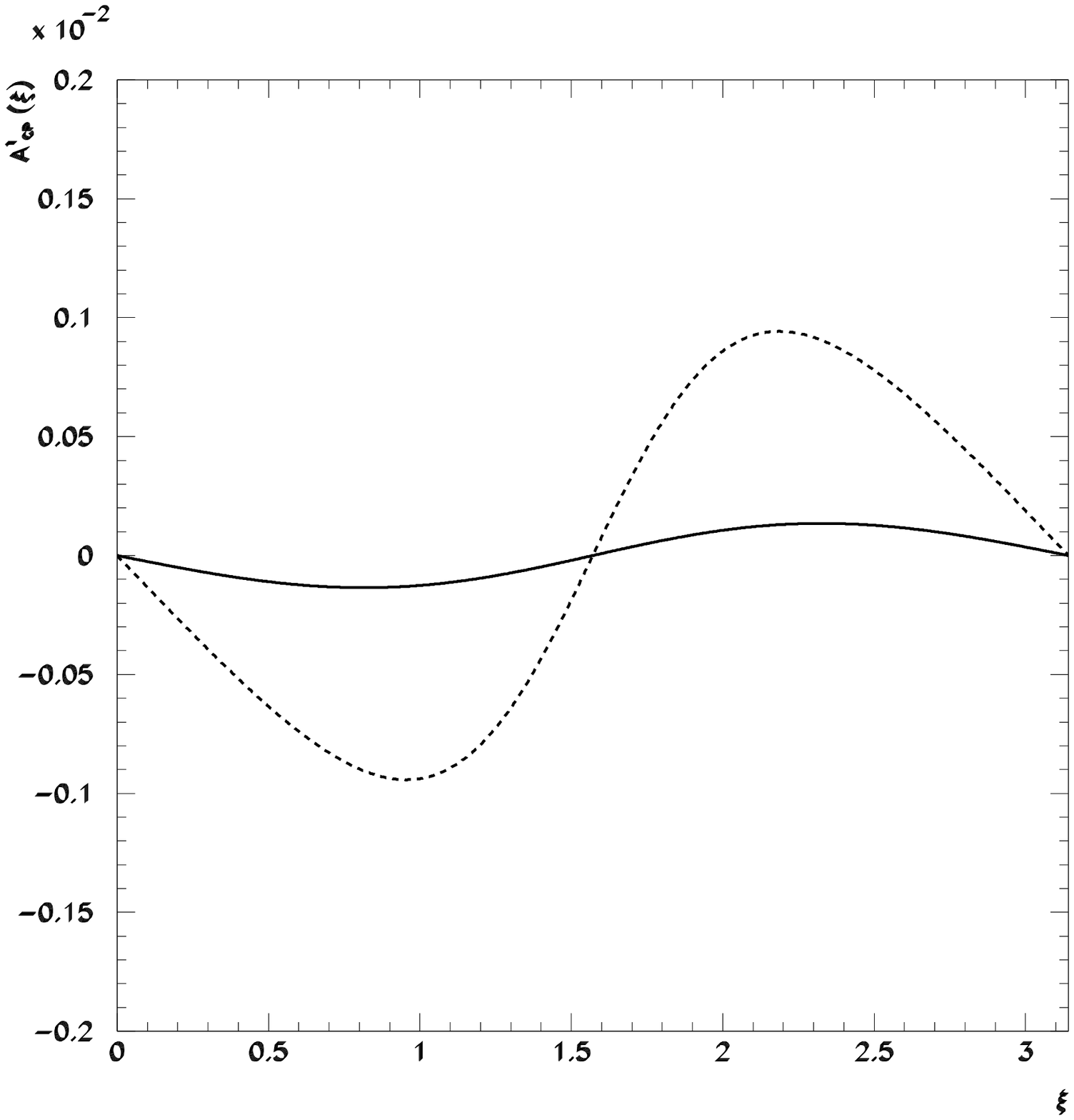}}
\caption[]{
$A^1_{CP}$ as function of $\xi$,
where $s=0.8$,  solid and dashed lines represent
$\tan\beta=10$ and $30$.}
\end{figure}


\begin{figure}
\epsfxsize=18 cm
\centerline{\epsffile{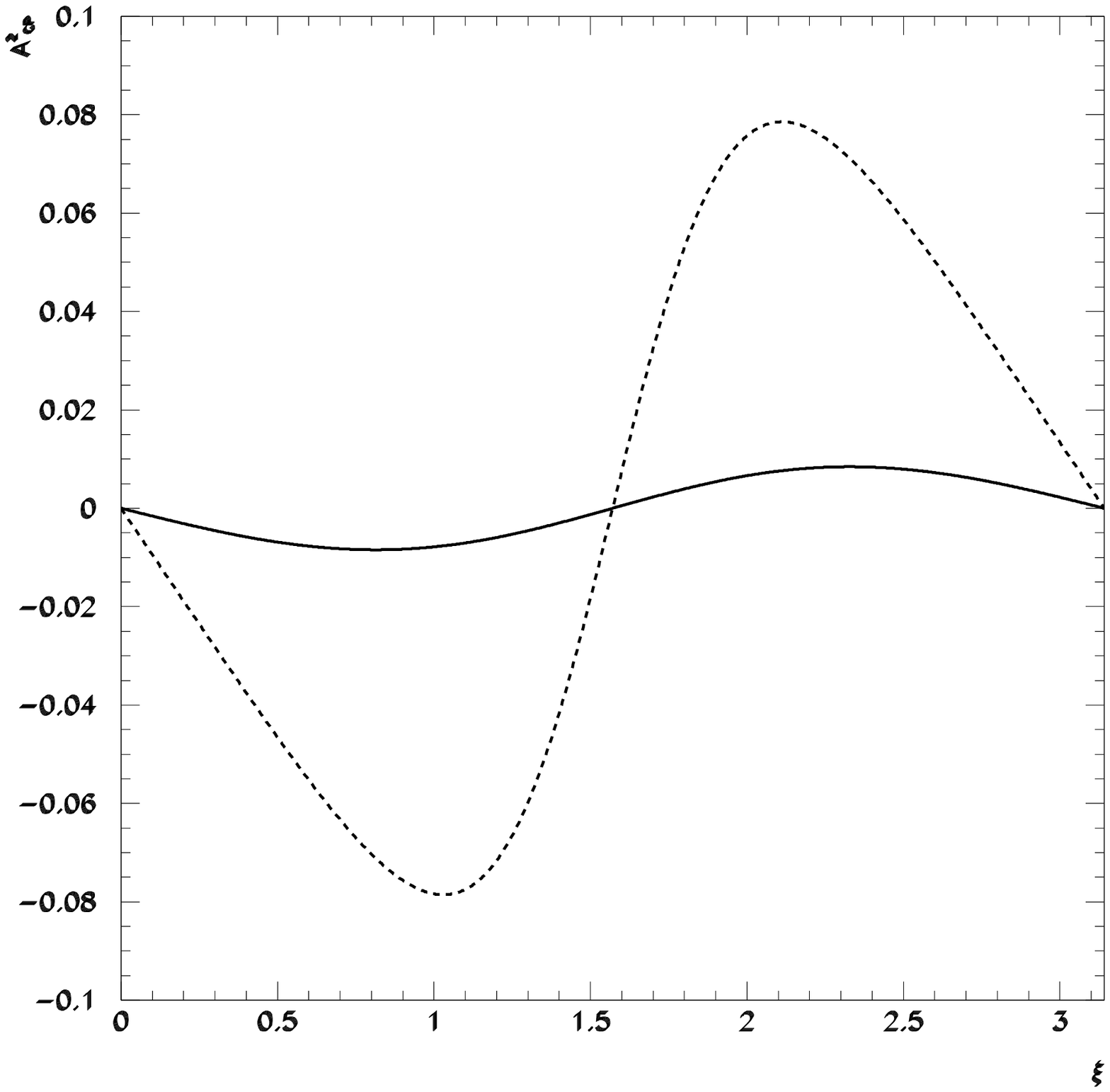}}
\caption[]{
$A^2_{CP}$ as function of $\xi$,
where $s=0.8$,  solid and dashed lines represent
$\tan\beta=10$ and $30$.}
\end{figure}

\begin{figure}
\epsfxsize=18 cm
\centerline{\epsffile{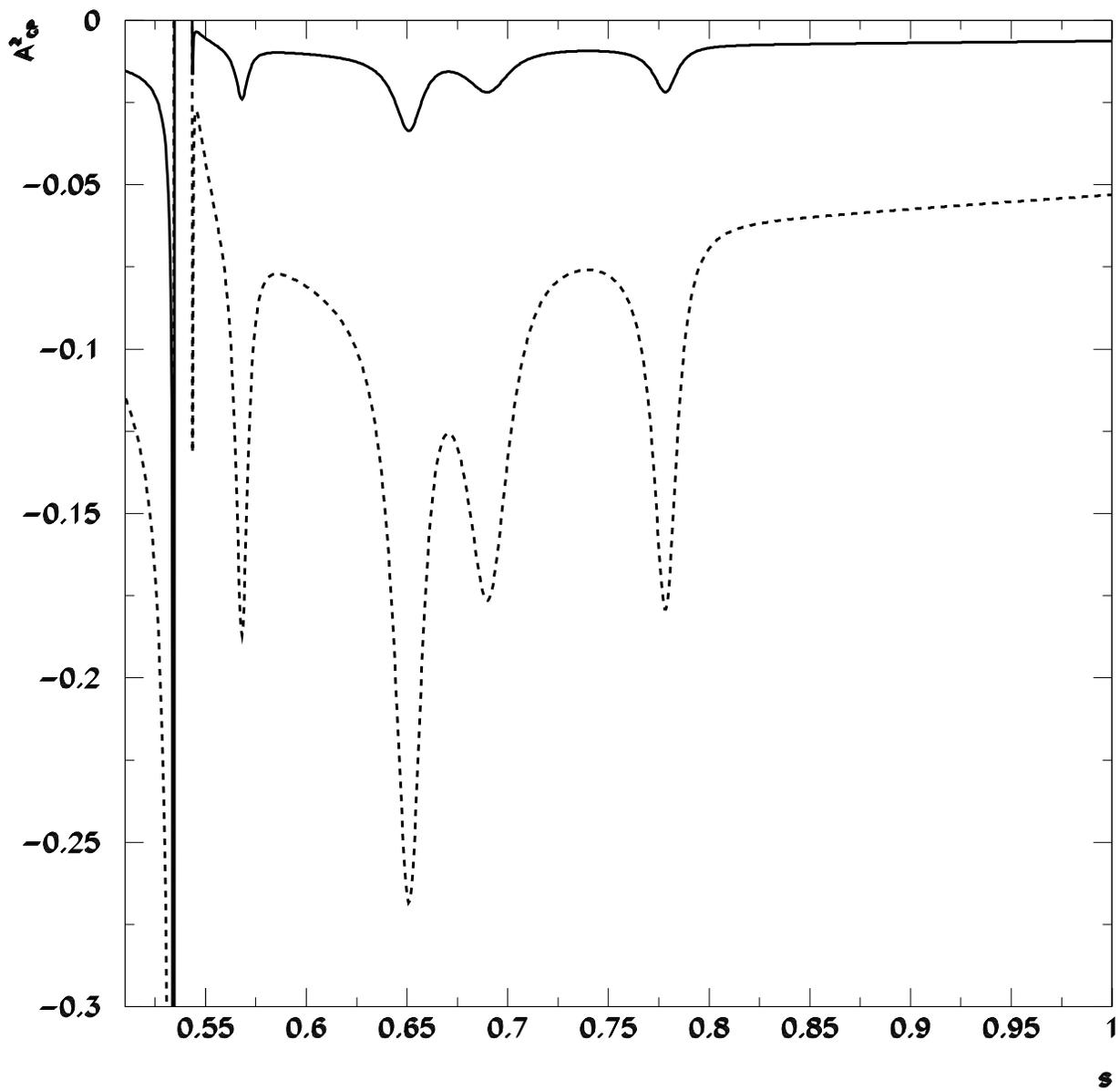}}
\caption[]{
$A^2_{CP}$ as function of $s$,
where $\xi=\pi/4$,  solid and dashed lines represent
$\tan\beta=10$ and $30$.}
\end{figure}


\begin{figure}
\epsfxsize=18 cm
\centerline{\epsffile{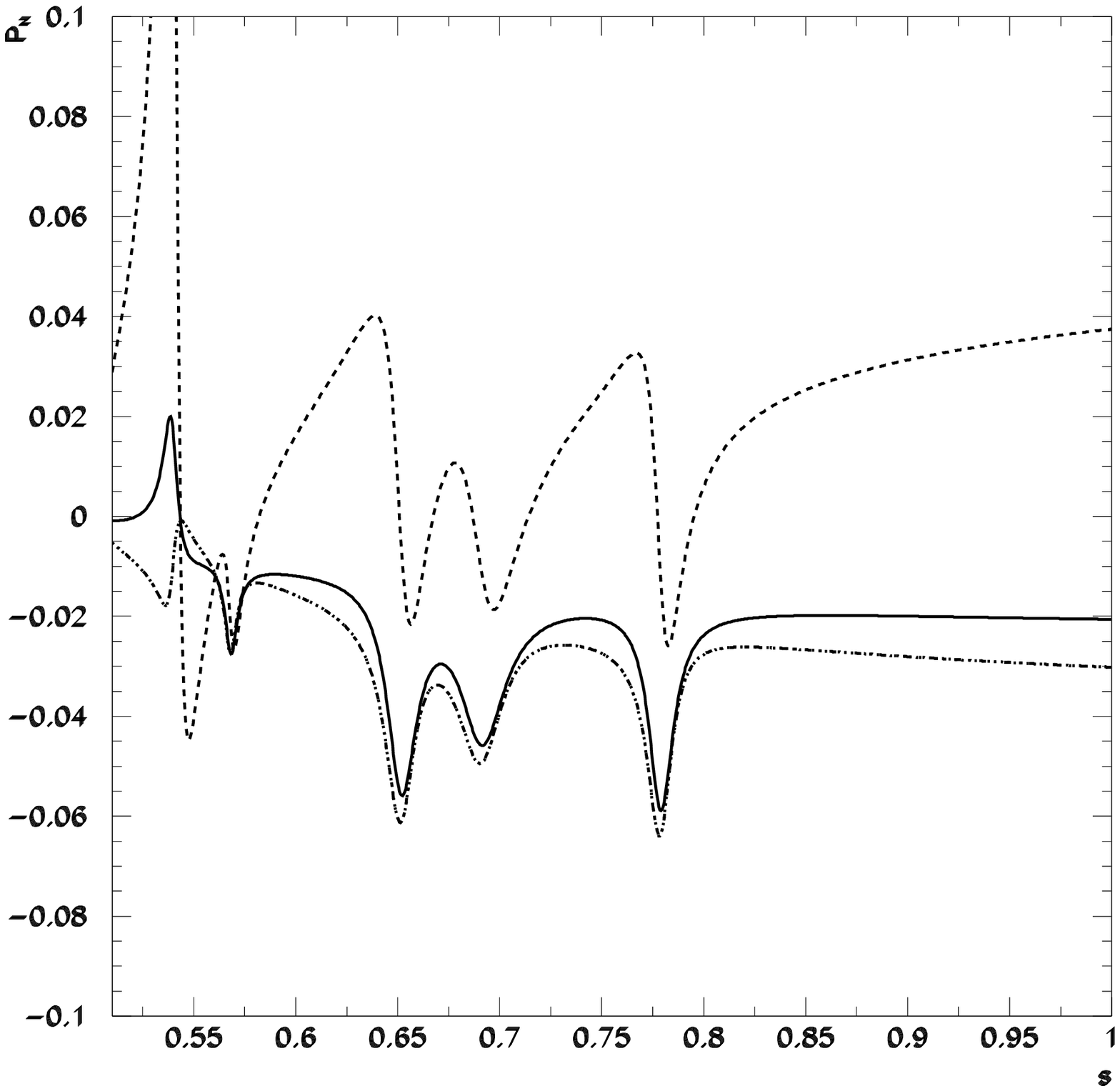}}
\caption[]{
$P_N$ as function of $s$,
where $\xi=\pi/4$,  solid and dashed lines represent
$\tan\beta=10$ and $30$, dot-dashed line represents the case of
switching off $C_{Q_i}$ contributions.}
\end{figure}

\begin{figure}
\epsfxsize=18 cm
\centerline{\epsffile{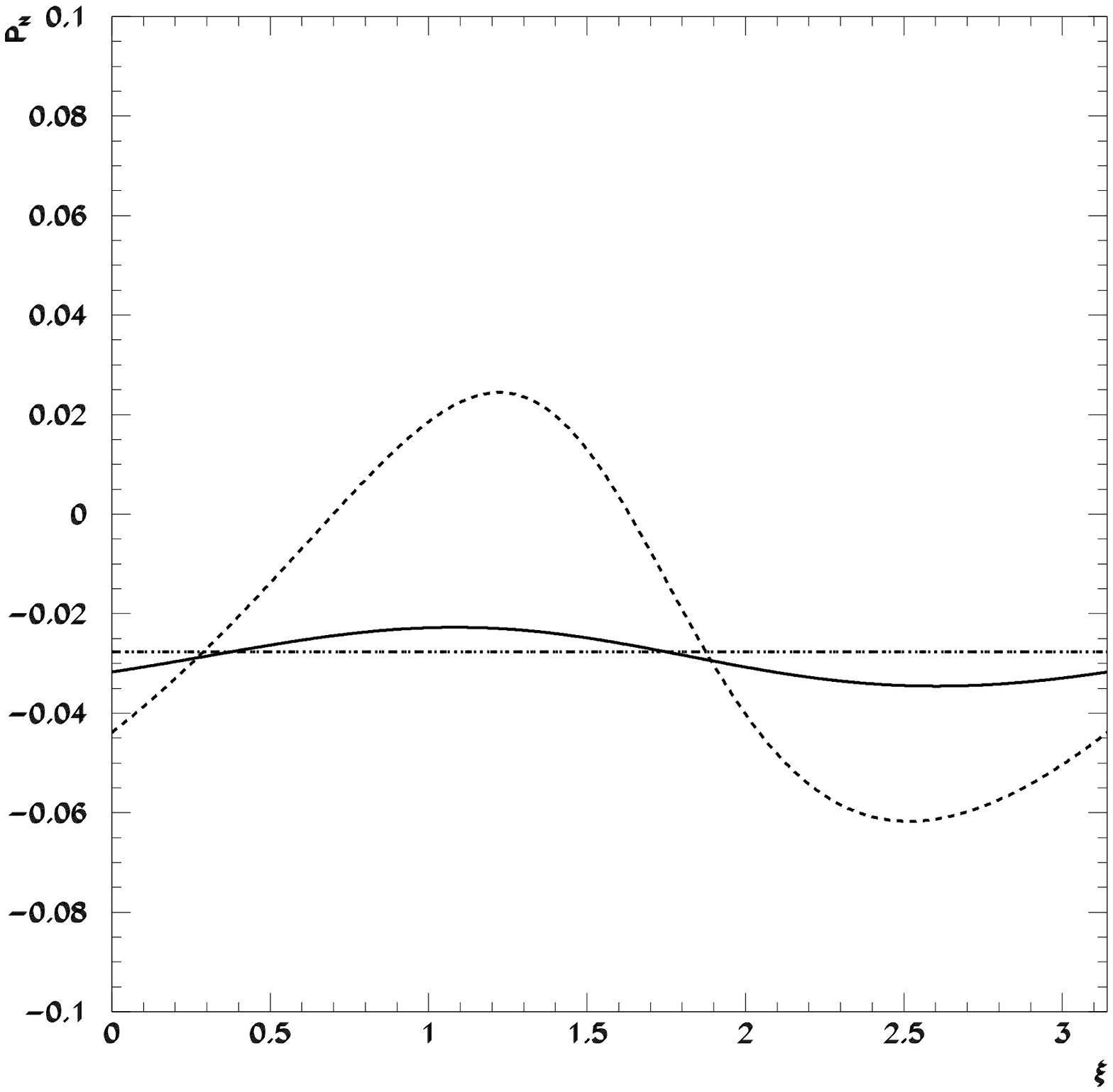}}
\caption[]{
$P_N$ as function of $\xi$,
where $s=0.8$,  solid and dashed lines represent
$\tan\beta=10$ and $30$, dot-dashed line represents the case of
switching off $C_{Q_i}$ contributions.}
\end{figure}


\begin{figure}
\epsfxsize=18 cm
\centerline{\epsffile{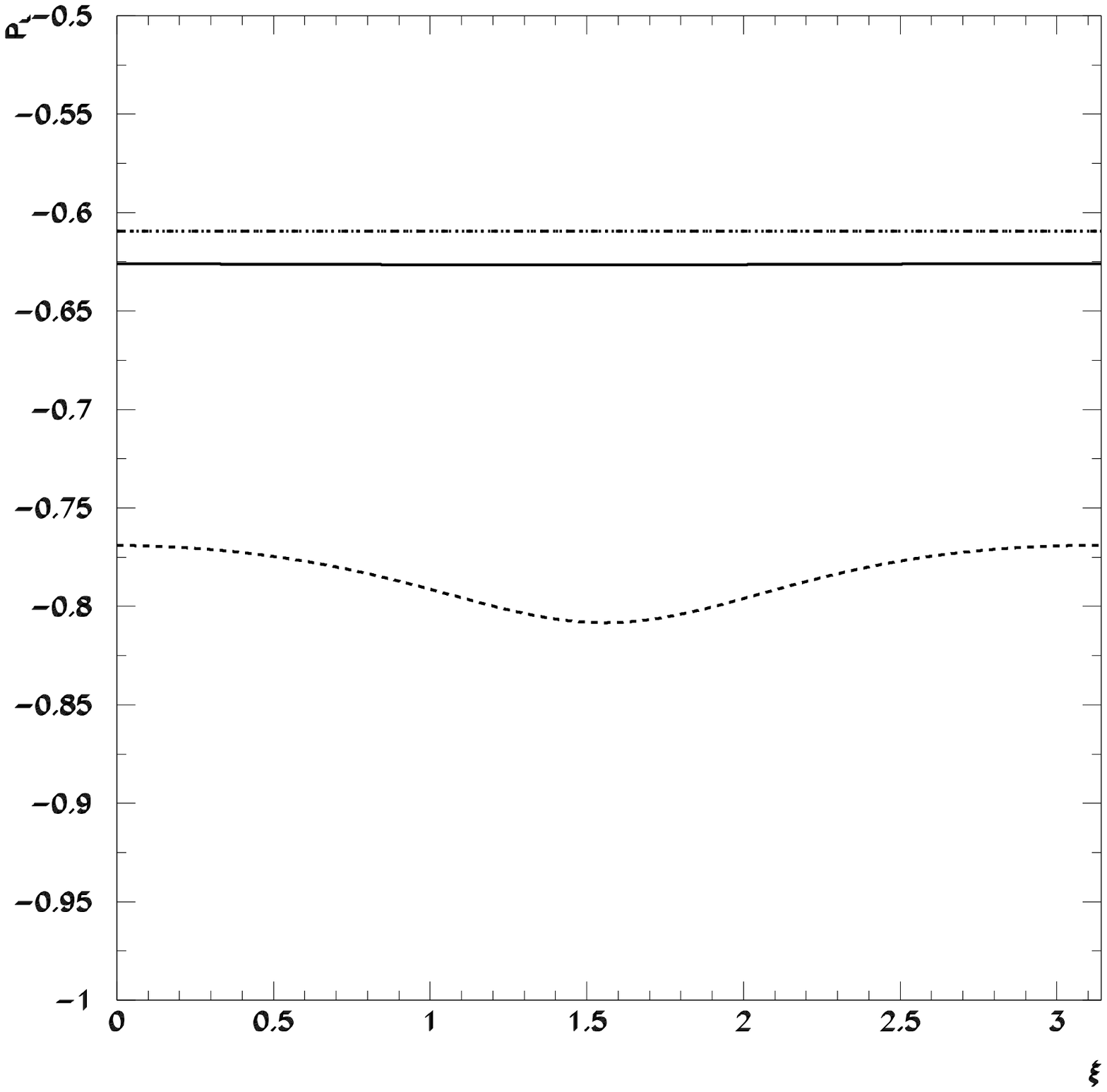}}
\caption[]{
$P_L$ as function of $\xi$,
where $s=0.8$,  solid and dashed lines represent
$\tan\beta=10$ and $30$, dot-dashed line represents the case of
switching off $C_{Q_i}$ contributions.}
\end{figure}

\begin{figure}
\epsfxsize=18 cm
\centerline{\epsffile{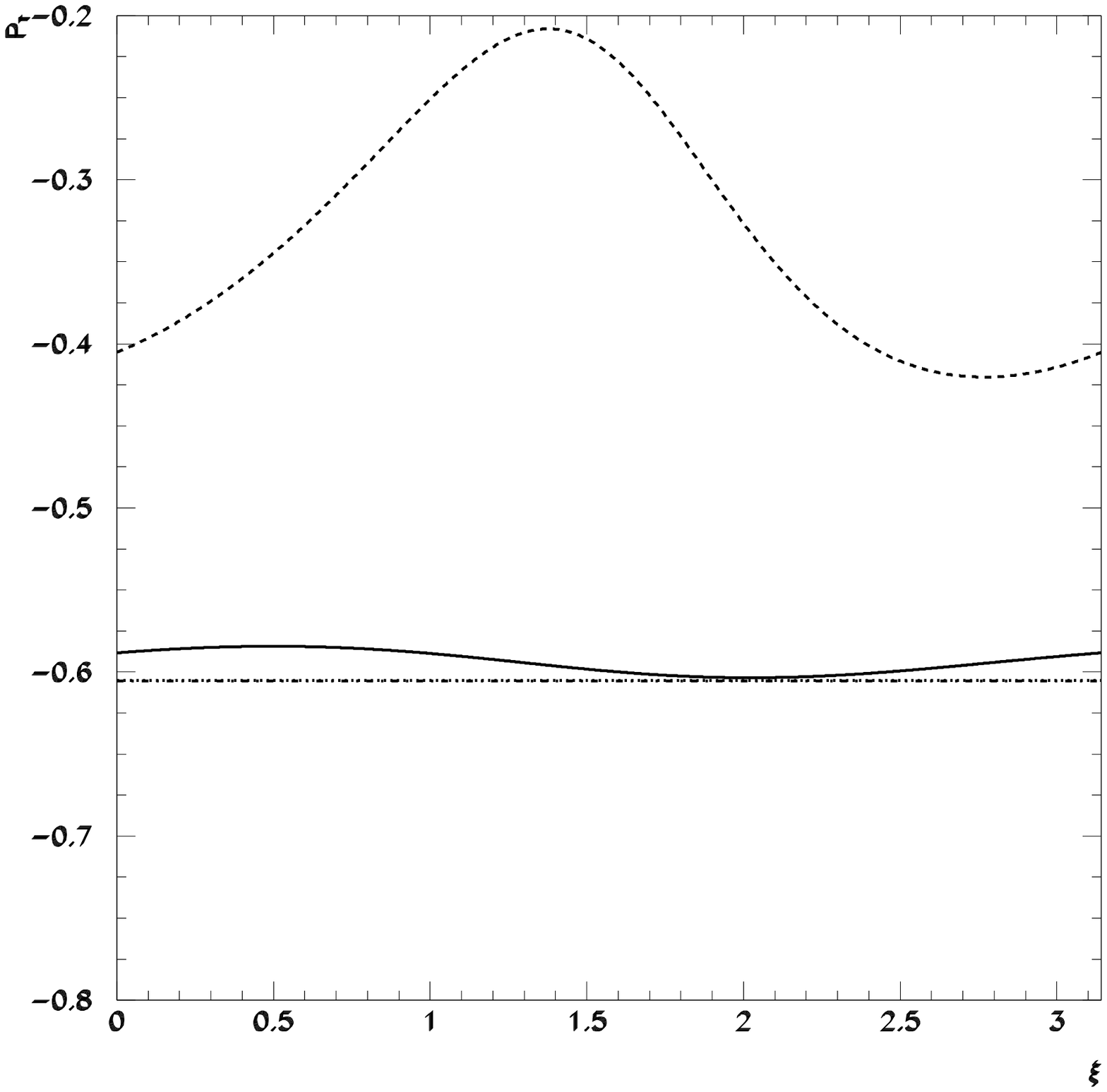}}
\caption[]{
$P_T$ as function of $\xi$,
where $s=0.8$,  solid and dashed lines represent
$\tan\beta=10$ and $30$, dot-dashed line represents the case of
switching off $C_{Q_i}$ contributions.}
\end{figure}


\end{document}